\documentclass[aps,prd,twocolumn,showpacs,nofootinbib,superscriptaddress]{revtex4-1}  

\usepackage{graphicx}  
\usepackage{bm}        
\usepackage{amssymb}   
\usepackage{hyperref}
\usepackage{natbib}
\usepackage{amsmath}
\usepackage{color}
\usepackage{ulem}

\def\simpropto{\lower.2ex\hbox{$\; \buildrel \propto \over \sim \;$}}
\def\ltsim{\lower.5ex\hbox{$\; \buildrel < \over \sim \;$}}
\def\gtsim{\lower.5ex\hbox{$\; \buildrel > \over \sim \;$}}

\begin{document}

\title{Molecular ionization rates and ultracompact dark matter minihalos}

\author {Joseph Silk}
\affiliation{Institut d'Astrophysique, UMR 7095 CNRS, Universit\'{e} Pierre et Marie Curie, 98bis Blvd Arago, 75014 Paris, France}
\affiliation{Department of Physics and Astronomy, The Johns Hopkins University, Homewood Campus, Baltimore MD 21218, USA}
\affiliation{Beecroft Institute of Particle Astrophysics and Cosmology, Department of Physics, University of Oxford, Oxford OX1 3RH}

\begin{abstract}
Molecular ionization in the Central Molecular Zone of our galaxy is enhanced over the typical galactic value by  an order of magnitude or more. This 
cannot be easily explained for dense Galactic Center  molecular complexes in the absence of embedded sources of low energy cosmic rays. We provide such a source in the form of ultracompact minihalos of self-annihilating dark matter for   a variety  of annihilation channels that  depend on the particle mass and model.  Such sources  are motivated for plausible inflationary power spectrum  parameters, and  while possibly subdominant in terms of the total dark matter mass within the Galactic bulge, might also account for, or at least not be in tension with,  the Fermi Galactic Centre $\gamma$-ray excess. 
 \end{abstract}
\maketitle

\section{Introduction\label{sec-intro}}

\smallskip
A large number of independent astronomical observations point towards the unexpected existence of  large numbers of molecular ions in dense clouds and correspondingly enhanced ionization
tracers in the Central Molecular Zone (CMZ), within 
the inner 0.3 kpc,  that contains much of the observed dense molecular gas. 
For example, methanol production requires a high cosmic ray ionization rate, $\sim 10^{-15}\rm s^{-1},$ in CMZ dark molecular clouds
\citep{2017MNRAS.467..737C}.
More extreme conditions are found in diffuse  near volume-filling molecular  clouds in the CMZ.
The measured ionization rate is $\zeta_H \sim 10^{-14} \rm s^{-1}$ in diffuse warm molecular clouds ($n\sim \rm 100cm^{-3} $, $T\sim 300\rm K$) \citep{2016A&A...585A.105L}.

In contrast, there is a track record of  somewhat lower but still unexpected ionization rates observed for molecular ions
in diffuse atomic ($OH^+$, $H_2 O^+$) and molecular clouds  ($H_3^+$) throughout the galactic disk  of order $\sim 10^{-16} \rm s^{-1}$
\citep {2017ApJ...845..163N}, 
with possible indications that these rates decrease with dust extinction.
However the directly measured disk ionization rate, inferred from Voyager 1 observations of low energy cosmic rays down to $\sim 3$ MeV\citep{2016ApJ...831...18C},  is an order of magnitude lower, around  $\sim 10^{-17}\rm s^{-1}$.

Other  potential indicators of enhanced  cosmic ray interactions within molecular gas   clouds  include detection of high CO rotational levels 
in starbursts
 \citep{2016A&A...591A..73G} and heating of molecular gas near AGN \citep{2018arXiv180802035L}.
The likely source of low energy cosmic rays is supernova remnant shocks \citep{2018arXiv180301794T}. 

The CMZ co-extends with the inner region of the Fermi diffuse gamma ray excess which continues out to about a kpc
\citep{2017ApJ...840...43A}.
{ 
Fermi provides a lower bound on
cosmic ray-induced ionization  via the cosmic ray hadronic interactions with dense gas. }
Simple estimates suggest that ionization rates are small for the standard spectrum of cosmic rays above a GeV, and well below those needed to account for ionization in the diffuse molecular interstellar medium \citep{2012MNRAS.425L..86N}. 
{
 A related phenomenon may be  the unexplained source of trace H atoms in dense molecular clouds \cite{2018arXiv180904168P}.} One needs to substantially augment 
the low energy cosmic ray spectrum within the  interiors of molecular clouds in order to produce enhanced ionization rates.

Early studies in less extreme situations postulated a low energy cosmic ray 
"carrot" in order to enhance the ionization in HI clouds  \citep{1968ApJ...152..971S} 
and spallation rates to contribute significantly to the observed  light element (Li, Be, B) abundances
 \citep{1971A&A....15..337M}. The recent 
diffuse interstellar  cloud ionization requirements 
are consistent with the 
associated
spallation arguments but cannot easily be reconciled with known cosmic ray sources \cite{2009ApJ...694..257I}.
Moreover dense clouds provide an even  more severe challenge to the carrot hypothesis. Cosmic rays are accelerated  by sources such as supernova remnants and diffuse throughout the interstellar medium. However their propagation is controlled by magnetic diffusion, which restricts their ability to penetrate dense interstellar clouds.  Low energy cosmic rays in particular are subject to ionization losses and they cannot effectively contribute to ionization or heating  in dense molecular clouds for any plausible magnetic field geometry  in the absence of local sources \citep{2018arXiv180410106P}.

Here I consider another source of low energy cosmic rays associated with dark matter in ultracompact minihalos (UCMHs).
 These 
 are generically predicted for non-scale-invariant primordial inflationary power spectra, with excess power on small scales that are not probed directly by CMB anisotropy observations \citep{2012PhRvD..85l5027B}. Such objects are likely to be a sub-dominant component of cold dark matter, and generate locally enhanced fluxes of dark matter (DM) annihilation products. UCMHs in the inner galaxy, especially in the CMZ, can contribute significantly to the molecular ionization rates without any excessive $\gamma$-ray signal.
Since many of the UCMHs coexist with, and are embedded in, the dense clouds,  they provide a local ionization source that is not restricted by magnetic shielding or ionization loss constraints. 

 \section {Ultracompact Minihalos and Dark Matter  in the CMZ}
 
{
The Galactic Center excess (GCE), measured by the  Fermi telescope to have  luminosity  $\sim 1-2\times 10^{37}\rm ergs /s,$
provided early support for self-annihilating CDM \citep{2016PDU....12....1D}, but has subsequently been found to be}
dominated by 1000s of weak gamma ray sources
\citep{2016PhRvL.116e1102B, 2016PhRvL.116e1103L, 2017ApJ...840...43A}. These sources 
are usually considered to be millisecond pulsars (MSPs).  In fact 
there is no strong observational evidence for, or against, a MSP interpretation \citep{2018PDU....20...88B}.
While the GCE data are inconsistent with an exclusively  diffuse  DM annihilation origin, UCMHs of annihilating DM provide a possible DM-dominated explanation, that would moreover fulfill the radial distribution constraint  of the GCE to be consistent with the bulge stellar distribution.  A similar case can be made for the diffuse gamma ray excess in M31
\citep{2018ApJ...862...79E}.  
{However here my focus is on a much softer annihilation signature of UCMHs  in the context of molecular ionization.}  We address UCMH survival issues below.

Constraints on the existence of such  low mass UCMHs at high redshift from $\gamma$-ray and CMB observations are highly constraining for gaussian initial conditions and especially steep intrinsic profiles \citep{2012PhRvD..85l5027B}.  While the shallow profiles favored by numerical simulations  weaken these constraints,
inclusion of  a broad mass spectrum actually strengthens them \citep{2018PhRvD..97b3539N,2018arXiv180607389S}. However inclusion of non-gaussianity in the initial conditions renders abundance calculations highly uncertain, cf. \cite{2018JCAP...03..016F}. 

Models are allowed with solar mass  UCMHs originating from fluctuations with enhanced power, of order 
$\delta P_k \sim1000$ times larger than at CMB angular anisotropy scales and  at $k\sim 10^{4.5}\rm Mpc^{-1}.$ The UCMHs can, in principle,  contribute a significant fraction of the DM.  Their epoch of formation  is of order $\delta P_k z_{gf},$ where $z_{gf} \sim 30$ is the epoch when nonlinearity first becomes important for the usual approximately scale-invariant fluctuation power spectrum. Non-gaussianity 
can further boost the initial UCMH fluctuation amplitude  to achieve nonlinearity at the earliest relevant epoch when DM structure can develop, namely  matter-radiation equality. 

 Clearly, there are many uncertainties in the initial conditions and structural properties of UCMHs, as well as 
 particle mass and associated annihilation channels. Rather than seeking to address these in detail,  with little prospect of convergence, I will proceed phenomenologically, and assume that UCMHs may constitute a substantial fraction $f_{UCMH}$ of the dark matter.
 To then derive annihilation rates,
 in the context of accounting for molecular ionization rates in the CMZ, I need to estimate the internal density within a UCMH,  and it is natural to normalize UCMH formation to the epoch of matter-radiation equality, when collapse can first occur.

The mass of DM in the CMZ is sensitive to the adopted DM profile. Observationally, this is only accessible by inwards extrapolation from the bulge/bar region, where
the dark matter content is constrained dynamically \citep{2017MNRAS.465.1621P}.
Within a  volume $\rm 29.6 \ kpc^3$,   the dynamical mass is $\rm 1.84 \times 10^{10}M_\odot$ and the dark matter mass (within 2 kpc) is  $5.10^9\rm M_\odot.$
 To extrapolate the dark matter  into the central 250 kpc is highly uncertain \citep{2017PDU....15...90I}: I assume two cases, a constant density core or a  NFW ($\propto 1/r$) profile.
 Extrapolation  for a NFW  cusp (favored for the GCE in the DM interpretation)
gives  a mass of  DM within  0.25 kpc  amounting to  $3.10^8 \rm M_\odot$. The conservative extrapolation is to a DM core in the CMZ, which contains about $10^7 M_\odot$  of dark matter.
 
The CMZ contains some 
10\% of the Milky Way Galaxy (MWG) molecular gas. 
{This amounts to $5 \pm 2 \times 10^7 \rm M_\odot$  of $H_2$  within the central 0.25 kpc \citep{2017ApJ...834...57M}.
The gas is mostly in 4  major GMC complexes (Sgr A,B,C,D) that are  almost contiguous, with typical density is $\rm 10^{4.2} cm^{-3}, $  and 
kinetic temperature  $10^{1.8}$ K. The gas is poorly correlated with star forming regions \citep{2018ApJS..236...40T}.}

 {We are interested in the ratio $ \xi= M_{DM}/M_{H2}$ in the central 0.25 kpc, or
 $\xi=5$ for a 1/r density profile;
 for  a core $M\propto r^3,$  the DM content  is reduced to give $\xi=0.2 $.
With increasing galactocentric distance, the numbers are more reliable: at 2 kpc, $M_{H_2}=2.4\times 10^8\rm M_\odot,$ $M_{DM}=5.10^{9}\rm M_\odot, $ and $\xi=20;$ at the solar circle, where the gas is predominantly atomic,
$M_{H_2}=2.10^9\rm M_\odot, $ $M_{DM}=7.10^{10}\rm M_\odot, $ and $\xi=35.$}

\section{Ultracompact minihalos and ionization rates}

I consider ionizations of nearby molecules by DM in UCMHs. In what follows, I  normalize the  dark particle mass $m_x$ to a GeV.
Since the UCMH profile rapidly evolves into a $\sim$$ r^{-1} -r^{-3/2}$  cusp \citep{2018PhRvD..97d1303D}
there is no significant cusp enhancement  of the annihilation flux and the  mean density in the UCMH  suffices  for computation of annihilation and associated ionization rates. 

The basic equation is the DM ionization rate determined by
(here $\zeta_H$  is ionization rate per H nucleus)
$\zeta_H n_{H2} = f_i\langle\sigma v\rangle n_x^2, Ê$ where $ f_i$
{
is the number of ionizations per WIMP annihilation, requiring 36 eV per ionization by energetic particles,  including the effects of secondary ionizations,  and $n_x=10^6 n_{x,6}\rm cm^{-3}
$ is the dark matter particle number density. I write
 $f_i= 10^{7.5}m^x_1 $ say for relativistic  protons and other products, eg in the MeV-GeV range, } and ignore
neutrino channels (which are $\sim$10\%).  Here $m^x_1=m_x/1\rm GeV.$

The internal density $n _x$ in UCMHs that formed as early as the epoch of matter-radiation equality is, at collapse,  
${3400}^3 170 n_{x,0}= 3.10^6  z_{3400}^3/m^x_1\rm \ cm^{-3},$ where $n_{x,0}=1.3. 10^{-6}\rm /m^x_1 \  cm^{-3}$. 
Here I parameterize  the equality epoch at redshift as 
$z_{3400} = (1+z)/3400$ 
\citep{2018arXiv180706209P}, but bear in mind that UCMHs could form at later epochs depending on the initial amplitude of the power spectrum
at small scales.

The resulting ionization rate is 
\begin{equation}
\zeta_H= Ê3.10^{-11}\rm s^{-1} f_{7.5} Ê \langle \sigma v\rangle_{26}  Ên_4^{-1}Ê
n_{x,6} ^2 /{m_x^1}.
\end{equation}
Here 
$\langle \sigma v\rangle_{26}= 10^{-26}\rm cm^{3}s^{-1} \langle\sigma v\rangle.$ and  $ n_{4}\equiv  \rm n_{H_2/}10^4cm^{-3}$ is the typical molecular  gas density in the CMZ.

This is large compared to the observed values in the most central CMZ molecular clouds,  of order $10^{-14}\rm s^{-1}, $
and in turn greatly exceeds the mean interstellar value $10^{-17}\rm s^{-1}$
in the atomic ISM. Of course this enhancement only affects gas in  the proximity of the UCMH, and I now evaluate the dilution factor due to the separation of the UCMHs. But I note already that 
the molecular gas  within some tens of UCMH radii is significantly  ionized.

The size of a UCMH is $R_{UCMH}=  \rm 10^{17}(M_1)^{1/3} z_{3400}^{-1}\rm  cm $ for a UCMH mass
$M_1=M_{UCMH}/M_\odot.$
The space density of UCMHs is $n_{UCMH}= 1.5 M_8 ( M_1 )^{-1}r_{250}^{-3}f_{UCMH} \rm pc^{-3}$ 
for a total DM mass in the CMZ of
$M_8=M_{CMZ}/10^8 \rm M_\odot$ 
and CMZ radius $r_{250}=r/250\rm  pc$, where $f_{UCMH}$ is the fraction of bulge dark matter in UCMHs.
I assume that the space distribution of DM in the bulge, for the DM in discrete clumps, follows that of the bulge stars, as also is consistent with the source of the Fermi excess  \citep{2017arXiv171104778B, 2018NatAs...2..387M}.

To obtain the ionization rate dilution factor, one needs to compare the UCMH size with the mean half-separation $\lambda$ between UCMHs
defined as $0.5 n_{UCMH}^{-1/3},$
 or  
$\lambda_{UCMH}=  M_8^{-1/3}  M_1^{1/3}  r_{250}f_{UCMH}^{-1/3} \rm pc$
 in the CMZ.
 The dilution factor is
 \begin{equation}
 f_{dil}= (R_{UCMH}/\lambda_{UCMH})^3=10^{-3.5}z_{3400}^{-3}M_8 r_{250}^{-3}f_{UCMH}.
 \end{equation}
 and is independent of UCMH mass.
  Hence one can tolerate a DM fraction of UCMH of $\sim 0.1$  for the previously derived value of $\zeta_H$ for a single UCMH. 
  One also needs to normalize to the total DM mass via the measured value of $\xi,$ the ratio of DM to  $H_2$ mass in the CMZ: with more enclosed DM, there are more ionizations per $H_2$ molecule.
  
  The volume-averaged ionization rate in the CMZ is 
  \begin{equation}\langle\zeta_H\rangle=
  Ê10^{-15}\rm s^{-1} \xi f_{7.5} Ê \langle \sigma v\rangle_{26}  Ên_4^{-1}ÊÊn_{x,6} ^2 
  z_{3400}^{-3}M_8 r_{250}^{-3}f_{UCMH} /m_x^1
  \end{equation}
 The implication is that even with a plausible value of the UCMH/dark matter fraction, chosen to avoid annihilation limits,  e.g. $f_{UCMH}\sim 0.1$,
 one can obtain large ionization enhancements  for a cuspy DM profile,
 with $\xi\sim 10.$   Diffuse molecular gas, at say $n_4\sim 0.01$,  especially in the central region of the CMZ, is likely to have an even higher  value of 
  $\zeta_H.$   
For those UCMHs which are embedded  in cloud cores, there is greatly reduced dilution. Hence the cores which account for $\sim 0.1\%$ of the CMZ molecular mass, albeit at a density of order $10^6 \rm cm ^{-3},$ could in some cases have even larger ionization rates.

\section{Comparison to Fermi Galactic Center excess}
I now demonstrate that the UCMH sources could contribute in $\gamma$-rays to the overall Fermi Galactic Center excess, and, if so, naturally  provide the observed fluctuations if a small fraction of the UCMHs are relatively massive.

We estimate the $\gamma$-ray flux normalized to the molecular ionization rate in the CMZ.  We take a mean ionization rate of $\rm 10^{-15}s^{-1} $ per H$_2$ molecule in the CMZ. With  the observed molecular mass $\rm 5.10^7 M_\odot$, we infer  an ionization rate 
 $\rm 4.10^{49} s^{-1}.$
There are   $10^{7.5 } m^x_1$  ionizations per annihilation, and so the annihilation  
rate is $1.4\times10^{42 }$ annihilations/s. 
The  $\gamma$-ray  luminosity is $\sim 1\%$ of the rate of annihilations  for GeV WIMPs. The excess luminosity associated with the CMZ ionizations is 
 $\sim 10^{40}$ GeV/s or $10^{37}$  ergs/s, comparable to the Fermi GCE.
                                                      
The annihilation time-scale  within the UCMH is $t_{ann}\sim (n_x\langle \sigma v \rangle )^{-1}= 5.10^{19}m^x_1z_{3400}^{-3}\langle \sigma v\rangle_{26}\rm \ s.$
The typical luminosity of a clump in annihilations is  
\begin{equation}
M_{UCMH}/t_{ann} \sim 3. 10^{34}\rm M_1 \langle \sigma v\rangle_{26}/m^x_1 \ ergs/s.
\end{equation}
Only a small fraction of this, of order 1\%, can be in prompt $\gamma$-rays, although the fraction is model,  and especially DM particle mass, dependent \citep{2011JCAP...03..051C}, and could be up to an order of magnitude smaller.

The contribution to the volume-averaged molecular ionization rate in the CMZ is independent of UCMH mass. This is also true for the volume
emissivity in $\gamma$-rays, which are of course constrained by the Fermi GCE. 
However massive UCMHs, while rare for typical initial conditions,  contribute large fluctuations. The expected UCMH mass spectrum is expected to be similar to that anticipated for primordial black holes \citep{2017PhRvD..96b3514C}.
One could have local hot spots with ionization 10 times the mean CMZ value, due to more massive UCMHs.
These would lead to locally enhanced ionization and  $\gamma$-ray fluctuations.

For the diffuse, nearly volume-filling H$_2$ at $\rm 100 \ cm^{-3} $ in the CMZ, the effect is potentially just as large. The lower molecular gas density compensates in part for the increased dilution of the UCMH annihilation sources.

\section{Discussion} 

The reason that UCMH sources are so effective  ionization sources is that the annihilation rate is locally boosted by some 6 orders of magnitude within each UCMH, whose density is known but whose mass is arbitrary, subject to observational constraints. Of course this is only important near the UCMHs. But these could collectively amount to 1-10\% of DM in the CMZ region, which would be equivalent to a significant fraction of the mass  of the CMZ molecular gas.
The parameter space is limited primarily by the dark mass fraction in the form of UCMHs.
Note that $\zeta_H$ cannot exceed the typical galactic value by more than two orders of magnitude or else  CO is destroyed \citep{2017ApJ...839...90B}.

I have not considered dynamical effects that could lead to destruction of the UCMHs in the central region of the galaxy. Contrary to the case made by 
\cite {2018JCAP...07..060C}, I believe that such arguments are inconclusive. Resolution issues in simulations show that they 
do not reliably account for dwarf galaxy abundances, cores or cusps. 
  The issue of the far denser UCMH softening and survival in the innermost galaxy is most likely beyond the reach of current simulations \citep{2018MNRAS.475.4066V}.  
  
  On the scales of interest here, the tidal disruption of UCMHs occurs at a galactic radius  where the enclosed density exceeds $n_x=
 n^x_\odot m^x_1\rm ( 10kpc/r)^\alpha, $  normalized to the DM density at the solar radius $n^x_\odot \sim 0.3 \rm GeV cm^{-3}$.
 This  tidal disruption comparison depends on the  inner density profile $\alpha$ within the solar radius.
 
 The value of $\alpha$ is highly uncertain, and of course is most likely a function of radius as for the usual mass profiles.
  For the DM mass within the CMZ radius  $\sim 0.25$ kpc, an estimate is that the enclosed dark mass extrapolation decreases as $M_{DM} \propto r^\alpha $ within 10 kpc  for $\alpha\approx 1$, giving  $10^8$ M$_\odot$ for NFW.  For a reduced scale length or a DM core, the enclosed DM mass is   less.
  The core scale is uncertain \citep{2013JCAP...07..016N}. More recent studies use bulge/bar stellar dynamical data  to  normalise to kpc scales, and may have a slight preference for a kpc scale  core
\citep{2017MNRAS.465.1621P}. In this case,
the enclosed dark matter mass  within the CMZ  central 0.5 kpc could be reduced by up to an order of magnitude, to $\sim 10^7 \rm M_\odot.$
 
 In either case, tidal disruption of UCMHs is unimportant. This is further vindicated by the fact that
 GMCs and globular clusters of comparable density exist within the CMZ. Tidal disruption should not be a major issue for UCMH survival in the CMZ or, equivalently, the region of the Fermi GCE. Of course, baryonic 
feedback complicates dwarf galaxy survival discussions, but   this is irrelevant for UCMH that are incapable of retaining enough cooling baryons to form stars.

In summary, the  rate of molecular (or atomic) ionizations due to a UCMH is 
\begin{equation}
\dot N^{UCMH}_{ioniz}=10^{45}z_{3400}^3\langle\sigma v \rangle_{26}M_1/{m^x_1} \rm  \ s^{-1}.
\end{equation}
This is diluted by the volume fraction occupied by UCMHs, $f_{dil}.$
The molecular ionization rate per unit volume is
$\langle\zeta_H\rangle=  \xi n_{UCMH} \dot N^{UCMH}_{ioniz} f_{ion},$
where 
$n_{UCMH} \propto f_{UCMH}$  
and
$\xi \sim 0.1 $ is the DM to molecular fraction in the CMZ, and is high  as $\sim 5$ for a cuspy DM profile
With characteristic values of  ionization fraction via annihilations $f_{ion}\sim 1, $ DM to molecule fraction $\xi\sim 10,$ and UCMH fraction of DM $f_{UCMH}\sim 0.1,$ 
  the ionization rate in the CMZ is of order $10^{-14}\rm s^{-1}$ per $H_2$ molecule for a cuspy DM profile.

The corresponding annihilation luminosity of a UCMH  is $3.10^{34} z_{3400}^3 \langle\sigma v \rangle_{26}M_1/ m^x_1 \rm ergs/s.$ Our model requires 
of order a million solar mass UCMHs within the CMZ for $f_{UCMH}\sim 0.1$  to account for the ionization rate,  but these must  not overproduce excessive  high energy $\gamma$-rays. For reasonable DM parameters that yield the required ionization rates in the CMZ, one matches, and hence provides a possible interpretation of,   the $\gamma$-ray luminosity of the GCE in the Fermi bands above a GeV, 
$\sim$$10^{37}\rm ergs/s.$ 

At lower energies,  hadronic cosmic ray interactions with the molecular gas  and other processes are important and dominate any putative GCE.
Hence only 1\% of the annihilation products is likely to emerge as $\gamma$-rays above a GeV. 
This is the primary constraint on the  proposed ionization model. One needs a source of ionizing particles that is relatively inefficient at producing $\gamma$-rays in the Fermi GCE range.

This should be well within the range of possible DM models. 
One might need to  appeal to subGeV dark matter particles, with the dominant annihilation channel via $e^+e^-$, in view of stronger $\gamma$-ray constraints  on an extended UCMH mass function in the case that  the DM  particle mass is above a GeV.

Source models, using low mass x-ray binaries, have been previously motivated \citep{2009MNRAS.392.1115B} by the INTEGRAL 511 keV flux from the bulge due to positron annihilations, mostly from the inner few degrees, of $\sim 2.10^{43}\rm s^{-1}$. While a dark matter annihilation model for the  INTEGRAL flux remains contrived because of spectral constraints on positron line emission \citep{2011RvMP...83.1001P}, similar models have been constructed for the GCE that are inefficient in producing high energy $\gamma$-rays 
and appeal to some $10^5$ weak $\gamma$-ray sources, such as   millisecond pulsars, in this region  in order to inject the required flux of MeV positrons \citep{2018MNRAS.tmp.2037B}. 

I conclude that if UCMHs constitute a few percent or more of DM and are not tidally destroyed in the dense inner regions of our galaxy, annihilations over a wide mass range of UCMHs could provide molecular ionization rates in the CMZ as large as $10^{-14}\rm s^{-1}.$ The extended nature of the UCMHs, implicit in their formation at matter-radiation equality,  provides an intriguing option for embedded sources of ionization in dense molecular clouds in the CMZ.  While $\gamma$-ray observations strongly constrain this model, spatial correlations should be sought between gamma ray fluxes and molecular ionization. The morphological similarity and extent of the CMZ to the region of the Fermi GCE suggests that  UCMHs could  provide the 1000s of $\gamma$-ray  sources  required by fluctuation analyses of the GCE, even if their  molecular ionization contribution is more limited. 

Compact HII regions are seen in the CMZ with no apparent ionization source, including 
 K$\alpha$  x-ray emission correlated with enhanced ionization of neutral gas 
\citep{2018MNRAS.473.2899P}.
There are also tens of dense dusty cores in the CMZ,  typically $\rm 100 -1000 M_\odot,$  with sizes $\sim$0.1 pc.  Of these, $\sim$80\% show no embedded star formation \citep{2018MNRAS.474.2373W}.
One possible ionization source  is  enhanced local ionization from rare but more massive UCMHs. Such objects  could act as seeds and undergo  substantial gas accretion for UCMH masses of order 1000
M$_\odot.$
 The Bondi accretion rate onto a 1000 solar mass UCMH is   $GM^2 n/v^3 =0.1M_{1000}^2n_4 v_{10}^{-3}\rm 
 M_\odot/yr,$ where  I have adopted a UCMH velocity of 10km/s. Substantial masses of molecular gas could be accreted by the most massive UCMHs.  Even in the presence of accreted gas, these
 UCMHs would have suppressed star formation, as the role of self-gravity due to the gas is reduced by the DM. 
 
 The modeling of the ionization rate is strongly constrained by the Fermi GCE. It is possible that enhanced ionization actually stimulates gas  cooling and dense core formation,  and one might thereby expect even the more massive UCMHs to have a molecular ionization signature in the CMZ. One should  seek evidence for a correlation between $\gamma$-ray fluctuations and molecular ions.
 Molecular observations  limit the masses and numbers of UCMHs if one assumes that the cosmic ray sources are embedded and hence  freely penetrate the dense clouds.
 
While I have focussed here on the CMZ, such signatures could play a role in the dark-matter dominated outer galaxy, where cool halo gas absorption studies of lensed SMM galaxies also show high ionization rates \citep{2018arXiv180804852I}. It is worth pointing that in the 3rd Fermi Source catalog, 30\% of sources, amounting to 1000 or so over 100 MeV-300 GeV, are unidentified \citep{2015ApJS..218...23A}. Many of these are potentially  in the outer galaxy and are possible massive ($\sim 1000 \rm M_\odot$) UCMH candidates. 
{Ionization rates in the local diffuse interstellar medium are also known to be enhanced over canonical cosmic ray-induced ionization rates. The UCMH dilution factor increases by $\sim 30$, whereas  the DM/gas  dominance 
increases by $\sim 3$ for a cuspy DM profile from  0.25 kpc  to 5-10 kpc.   Ionization induced by DM annihilations might play a role at galactic radii where DM is more dominant, with ionization rates an order of magnitude lower than in the CMZ.
 }

The constraints on the role  of UCMHs as dark matter components, which  constrain  their mass range,  formation epoch and central density enhancement, are highly uncertain.
This leaves open the role of other  ionization sources. A more conventional astrophysical source in the CMZ might be  a  sufficiently large population  of faint low mass x-ray binaries or millisecond pulsars, as advocated in \cite{2018JCAP...07..042G}.
Primordial  black holes in the 
20-100 M$_\odot$ mass range (motivated by LIGO 
results \citep{2016PhRvL.116t1301B})  also remain a viable  but subdominant dark matter component \citep{2017PhRvD..96l3523A},   with accretion onto black holes  being  enhanced in dense molecular clouds and possibly providing  a comparable embedded source of ionization.  
 
\label{sec-Con}

\begin{acknowledgments}
I thank Mathieu Boudaud, David Neufeld, Tomohiro Nakama and Martin Stref for useful discussions.
\end{acknowledgments}


\end{document}